\documentclass[12pt,a4paper,twoside]{article}
\usepackage{epsfig}

\long\def\ifundef#1#2#3{\expandafter\ifx\csname
  #1\endcsname\relax#2\else#3\fi}
\ifundef{ensuremath}{\newcommand{\ensuremath}[1] {\relax\ifmmode {#1}\else {$#1$} \fi}}{}
\ifundef{mathcal}{\newcommand{\mathcal}[1]{{\cal #1}}}{}
\ifundef{mathrm}{\newcommand{\mathrm}[1]{\rm #1}}{}
\ifundef{resizebox}{\newcommand{\resizebox}[3] 
{#3 
}}{}
\ifundef{includegraphics}{\newcommand{\includegraphics}[1] 
{$#1$ 
\typeout{PostScript File #1 not included!}
}}{}

\newcommand{\epem}   {\ensuremath{\mathrm{e^+e^-}}}

\newcommand{\as}     {\ensuremath{\alpha_s}}

\newcommand{\asq}    {\ensuremath{\alpha_s(Q)}}
\newcommand{\asmz}   {\ensuremath{\alpha_s(M_{\mathrm{Z^0}})}}

\newcommand{\oaa}    {\ensuremath{\mathcal{O}(\alpha_s^2)}}

\newcommand{\znull}  {\ensuremath{\mathrm{Z^0}}}

\newcommand{\mz}     {\ensuremath{M_{\mathrm{Z^0}}}}

\newcommand{\chisq}  {\ensuremath{\chi^2}}
\newcommand{\chisqd} {\ensuremath{\chi^2/\mathrm{d.o.f.}}}
\newcommand{\xmu}    {\ensuremath{x_{\mu}}}

\ifundef{pt}{ }
            { }

\newcounter{hours}
\newcounter{minutes}

\newcommand{\Printtime}{%
  \setcounter{hours}{\time/60}%
  \setcounter{minutes}{\time-\value{hours}*60}%
  \ifthenelse{\value{hours}<10}{0}{}\thehours:%
  \ifthenelse{\value{minutes}<10}{0}{}\theminutes}

\begin{document}
\bibliographystyle{utphys}
%
\begin{titlepage}
\vspace*{-10mm}
\hbox to \textwidth{ \hsize=\textwidth
\hspace*{0pt\hfill} 
\vbox{ \hsize=58mm
{
}
}
}

\bigskip\bigskip\bigskip\bigskip\bigskip\bigskip\bigskip\bigskip\bigskip
\begin{center}
{\Huge\bf
{\ensuremath{C}}-Parameter and Jet Broadening \\[2.5mm]
at PETRA Energies
}

\end{center}
\bigskip\bigskip
\begin{center}
O.~Biebel$^{(1)}$, 
P.A.~Movilla~Fern\'andez$^{(1)}$, 
S.~Bethke$^{(1)}$ \\
 and the JADE Collaboration$^{(2)}$
\end{center}
\bigskip

\begin{abstract}
\noindent 
 e$^+$e$^-$ annihilation data recorded by the JADE detector
 at PETRA were used to measure the $C$-parameter for the
 first time at $\sqrt{s}= 35$ and $44$ GeV. The distributions
 were compared to a resummed QCD calculation.
 In addition, we applied extended resummed calculations to 
 the total and wide jet broadening variables, $B_T$ and $B_W$.
 We combined the results on $\alpha_s$ with those of our previous
 study of differential $2$-jet rate, thrust, and heavy jet mass,
 obtaining
 $\alpha_s(35 {\mathrm{~GeV}}) = 0.1448 ^{+0.0117}_{-0.0070}$
 and
 $\alpha_s(44 {\mathrm{~GeV}}) = 0.1392 ^{+0.0105}_{-0.0074}$.
 Moreover power corrections to the mean values of the observables 
 mentioned above were investigated considering the Milan factor and 
 the improved prediction for the jet broadening observables. Our study,
 which considered e$^+$e$^-$ data of five event shape observables 
 between $\sqrt{s}= 14$ and $183$ GeV,
 yielded $\alpha_s(M_{\mathrm{Z^0}})=0.1177 ^{+0.0035}_{-0.0034}$.
\end{abstract}

\vspace*{0pt\vfill}
\vfill
\bigskip\bigskip\bigskip\bigskip

{
\small
\noindent
$^{(1)}$ 
\begin{minipage}[t]{155mm} 
III. Physikalisches Institut der RWTH Aachen,
D-52056 Aachen, Germany \\
contact e-mail: Otmar.Biebel@Physik.RWTH-Aachen.DE
\end{minipage}\\
$^{(2)}$ 
\begin{minipage}[t]{155mm} 
for a full list of members of the JADE Collaboration see Reference \cite{bib-naroska}
\end{minipage}
\hspace*{0pt\hfill}
}

\end{titlepage}
%
%
\newpage
\section{ Introduction }

In a recent publication~\cite{bib-newJADE} we have presented a study
on event shape variables and determinations of \as\ using data of
\epem\ annihilations at $\sqrt{s} = 22$ to $44$~GeV recorded with the
JADE detector \cite{bib-JADEdet} at the PETRA collider.  This
study provided valuable information which was not available before
from $\epem$-annihilations in the PETRA energy range. The results on
\as, obtained in a manner similar to the determinations at
LEP, demonstrated that the energy dependence of \asq\ is in good
agreement with the prediction of Quantum Chromodynamics (QCD). Evolved
to the $\znull$ mass scale, the results agree well with
those obtained at LEP, and are of similar precision. In addition,
power corrections, applied to analytic QCD calculations of the mean
values of event shape distributions, were found to qualitatively and
quantitatively describe the effects of hadronisation. Thus QCD could 
be tested without the need of phenomenological hadronisation models.

Meanwhile the perturbative calculations for the jet broadening
variables were improved by including a proper treatment of the quark
recoil~\cite{bib-new-jetbroadening}.
Furthermore, a resummation of leading and next-to-leading logarithm 
terms to all orders of \as\ (NLLA) became available for the 
$C$-parameter~\cite{bib-C-resummation}.
Besides these advances in the perturbative description of event shape
observables progress was made in the understanding of non-perturbative
power corrections to the event shape observables and their mean
values. In particular two-loop calculations of such corrections were
performed which modify the one-loop result of the power correction to 
the event shapes by a factor (Milan factor) \cite{bib-Milan-factor}.
Revisiting the interdependence between perturbative and non-perturbative 
contributions for the case of jet broadening observables yielded in 
Reference~\cite{bib-new-broadening} a further modification of the results 
given in~\cite{bib-new-jetbroadening}.

In this paper we complement our previous publication \cite{bib-newJADE} 
by a new \oaa+NLLA determination of \as\ from $C$-parameter and update our
\as\ determination from jet broadening at $\sqrt{s} = 35$ and $44$ GeV in
the Sections~\ref{sec-procedure} and \ref{sec-alphas}.  We also
applied power corrections to the $C$-parameter distributions and
re-investigated those for the mean values of the thrust, heavy jet
mass and both jet broadening observables in
Section~\ref{sec-meanvalues}. We start in Section~\ref{sec-data} with
a brief summary of the data samples used and draw conclusions from our
results in Section~\ref{sec-conclusions}.

\section{ JADE data }
\label{sec-data}

The JADE detector was operated from 1979 until 1986 at the PETRA
electron-positron collider at centre-of-mass energies of $\sqrt{s} =
12$ to $46.7$ GeV.  A detailed description of the JADE detector can be
found in~\cite{bib-naroska,bib-JADEdet}.  The components mainly used
for this analysis were the central jet chamber to measure charged particle
tracks and the lead glass calorimeter to measure energy depositions of
electromagnetic showers, which both covered almost the whole solid
angle of $4\pi$.

For the present study we consider data collected by the JADE detector
in 1984 to 1986 at centre-of-mass energies between $39.5$-$46.7$~GeV
and at $35$~GeV.  Multihadronic events were selected by the standard
JADE selection cuts~\cite{bib-JADEtrigger} which have been summarised in
detail in our previous publication~\cite{bib-newJADE}. Background from
two-photon processes and $\tau$-pair events and from events with hard
initial state photon radiation were mainly suppressed by posing
constraints on the visible energy, the total momentum and the charged
particle multiplicity of the event, and by cuts in the polar
angle of the thrust axis. The estimated purity of the resulting data
samples is about 98\% \cite{bib-JADEeventsel}. The numbers of events
retained after these cuts are $6158$ around $\sqrt{s} =$ $44$~GeV 
and $20\thinspace 926$ at $\sqrt{s} =$ $35$~GeV.

Corresponding {\em original} Monte Carlo detector simulation data for
$35$ and $44$~GeV were based on the QCD parton shower plus string
fragmentation implemented by the JETSET 6.3 event generator~\cite{bib-JETSET}.  
Both
samples included a detailed simulation of the acceptance and resolution 
of the JADE detector.  We used the standard set of parameters for the 
event generation as described in~\cite{bib-newJADE}. As pointed out
there, the agreement of the simulated data with the measurements is
good.

\section{ Measurement of event shapes}
\label{sec-procedure}

From the data passing the selection criteria as described in the
previous section, the distributions of the $C$-parameter and the 
jet broadening event shape variables $B_T$ and $B_W$ were determined 
using the 3-momenta $\vec{p}_i$ of the reconstructed particles.
The $C$-parameter is defined in Reference~\cite{bib-C-parameter}
as
\begin{displaymath}
            C = 3 (\lambda_1 \lambda_2  + \lambda_2 \lambda_3  + \lambda_3 \lambda_1 )
\end{displaymath}
where $\lambda_\gamma$, $\gamma=1, 2, 3$, are the eigenvalues of the momentum tensor
\begin{displaymath}
     \Theta^{\alpha\beta} = \frac{\sum_i \vec{p}_i^{\,\alpha} \vec{p}_i^{\,\beta} / |\vec{p}_i|}
                                 {\sum_j |\vec{p}_j|}
   \ \ \ .
\end{displaymath}
The definition of the total and wide jet broadening measures can
be found in Reference~\cite{bib-NLLA-1}.

Limits of the acceptance and resolution of the detector and effects
due to initial state photon radiation were corrected by a
bin-by-bin correction procedure.  The
correction factors were defined by the ratio of the distribution
calculated from events generated by JETSET 6.3 at {\em hadron level}
over the same distribution at {\em detector level}.  The {\em hadron
level} distributions were obtained from JETSET 6.3 generator runs
without detector simulation and without initial state radiation, using
all particles with lifetimes $\tau > 3\cdot 10^{-10}$~s.  Events at
{\em detector level} contained initial state photon radiation and a
detailed simulation of the detector response, and were processed in
the same way as the data.

\begin{figure}[!htb]
\begin{center}
\resizebox{79mm}{!}{\includegraphics{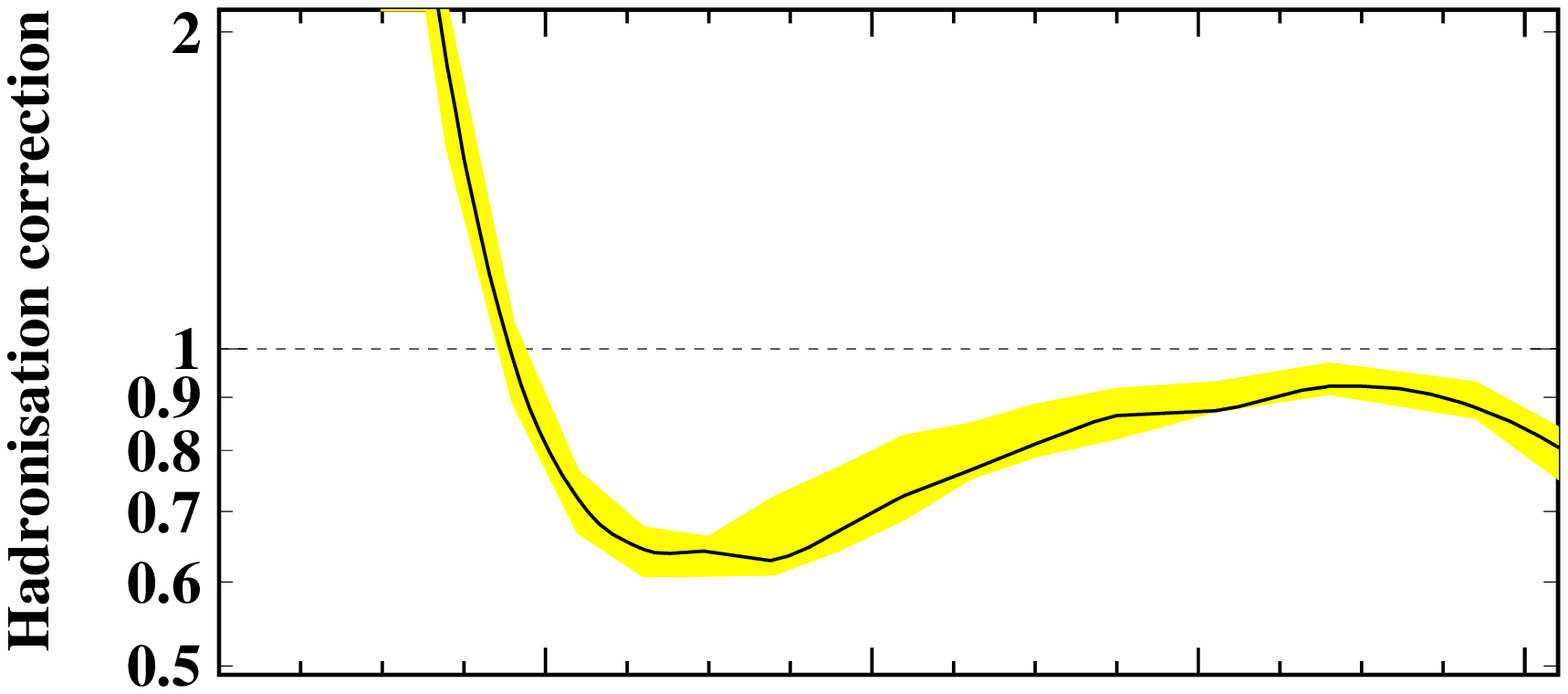}}
\resizebox{79mm}{!}{\includegraphics{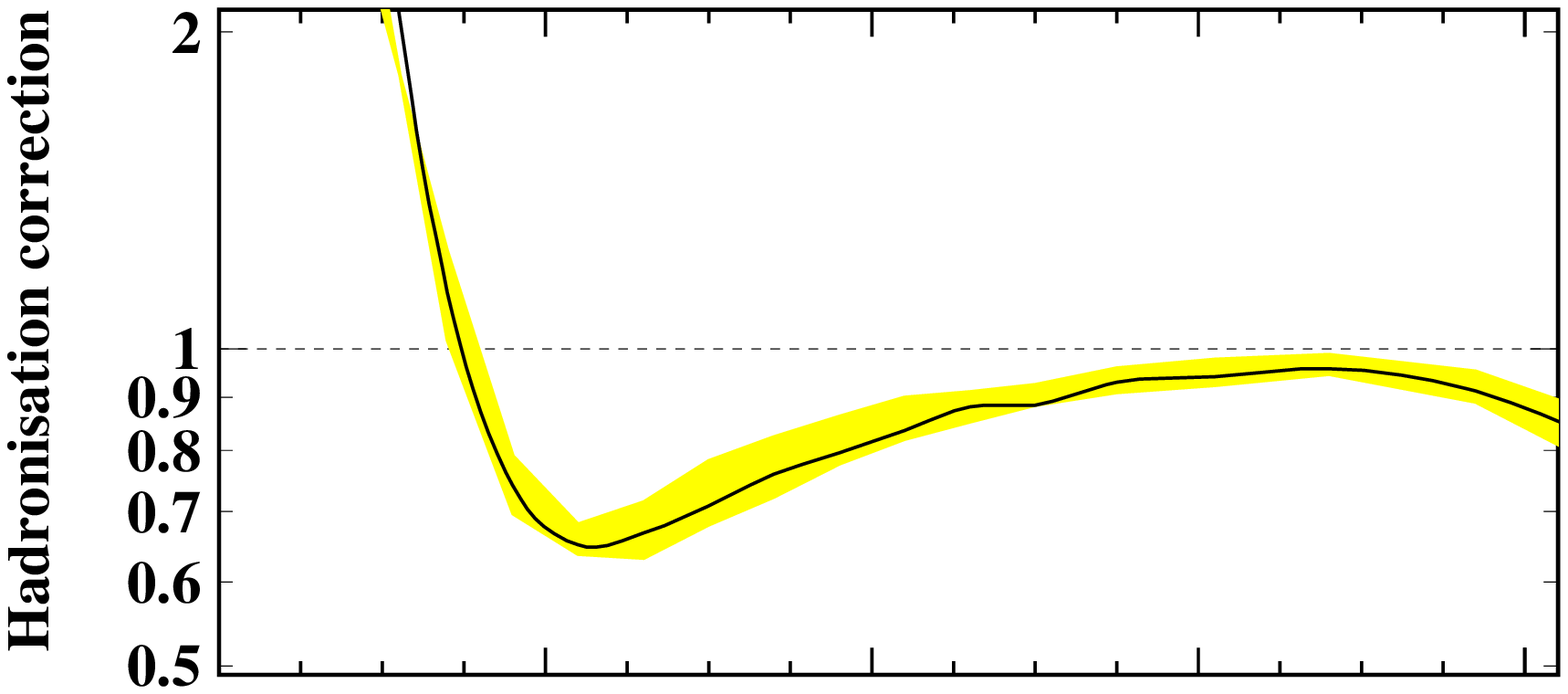}} \\[-18.6mm]
\resizebox{79mm}{!}{\includegraphics{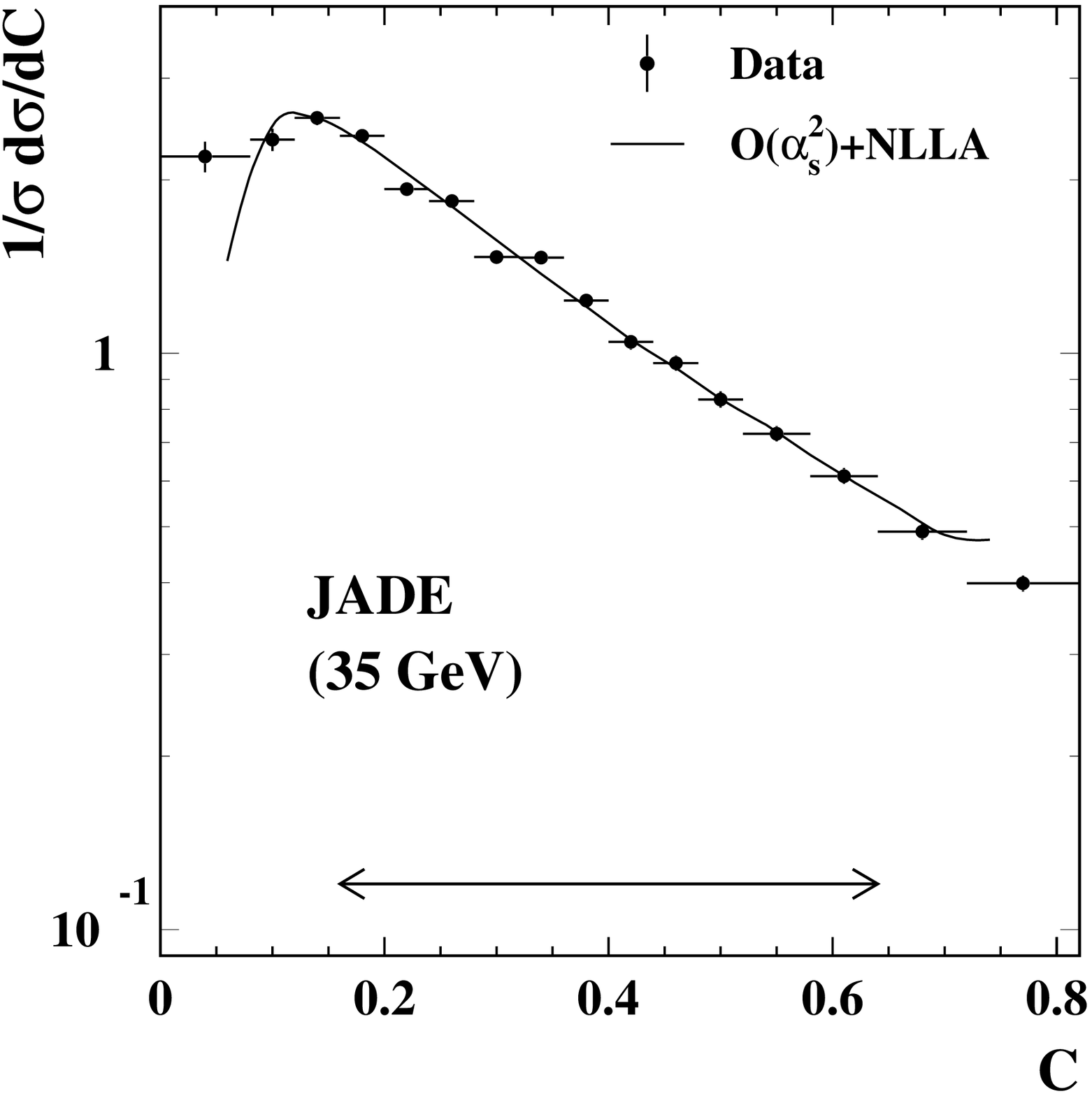}}
\resizebox{79mm}{!}{\includegraphics{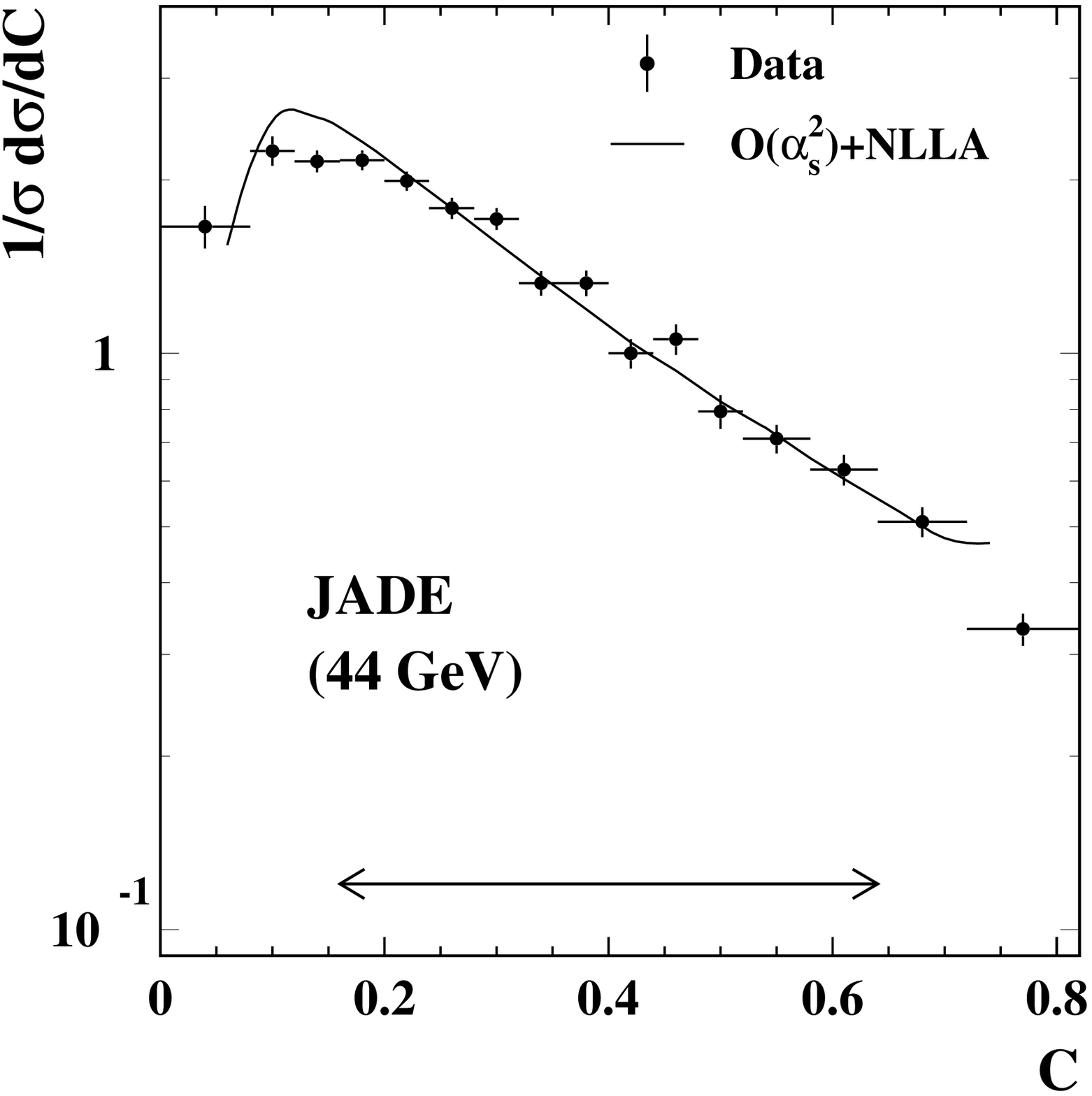}}
\end{center}
\caption{\label{fig-asresult}
  Distributions for $C$-parameter (bottom) measured at
  $\protect\sqrt{s} = 35$ and $44$~GeV and corrected to {\em parton
    level}.  The fits of the new \oaa+NLLA QCD calculation for the
  $C$-parameter are overlaid and the fit ranges are indicated by the
  arrows.  The error bars represent statistical errors only. Also
  shown are the hadronisation correction factors (top). The shaded
  bands represent the uncertainties due to the modelling of the parton
  shower and string fragmention by JETSET 6.3.}
\end{figure}

The data distributions were further corrected for hadronisation
effects by applying bin-by-bin correction factors derived from the
distributions at {\em parton level} and at {\em hadron level}, where
the parton level is given by the partons before hadronisation.
In the upper part of Fig.~\ref{fig-asresult}, 
the hadronisation correction for the distributions of the $C$-parameter at
$\sqrt{s} = 35$ and $44$ GeV are shown.  The correction factors for
all event shapes are typically of the order $10$ to $30$\% increasing
towards the $2$-jet region.

Systematic uncertainties of the corrected data distributions were
investigated by modifying details of the event selection and of the
correction procedure. For each variation the whole analysis was
repeated and any deviation from the main result was considered as a
systematic error.  In general, the maximum deviation from the main
result for each kind of variation was regarded as a symmetric systematic
uncertainty.  The main result was obtained using the default selection
and correction procedure as described above.

To estimate experimental uncertainties, we considered either tracks or
clusters for the measurement of the event shape distributions.
Furthermore, the selection cuts for multihadronic events were either
loosened or tightened as described in~\cite{bib-newJADE}, in order to
check the influence of background events on the measurement.

The impact of the hadronisation model of the JETSET 6.3 generator was
studied by varying the values of several significant model parameters
around their tuned values from Reference~\cite{bib-JADEtune} used for
our main result.  The variations amount to the percentage of the one 
standard deviations found in Reference~\cite{bib-OPALtune} from a 
parameter tuning of JETSET.  In detail, effects due to parton shower, 
hadronisation parameters, and quark masses were considered, following 
the procedure in~\cite{bib-newJADE}. These
uncertainties of the hadronisation corrections are shown for the
$C$-parameter distributions at $\sqrt{s} = 35$ and $44$~GeV as shaded
band around the correction factors in the upper part of
Figure~\ref{fig-asresult}.

\section{Determination of \as\ at $\sqrt{s} = 35$ and $44$ GeV}
\label{sec-alphas}
The data values for the event shape distributions and the
corresponding mean value of the $C$-parameter
corrected to {\em hadron level} are listed in
Table~\ref{tab-eventshapes-35+44GeV}, where also the statistical
errors and experimental systematic uncertainties are given. 
Our measurements of the jet broadening event shape distributions can
be found in Reference~\cite{bib-newJADE}.

After correcting for hadronisation, the event
shape distributions were compared directly with the analytic QCD
calculations. We determined the strong coupling constant \as\ by
\chisq\ fits to the event shape distributions of $C$, $B_T$, and
$B_W$. 
The fits to the
$C$-parameter distributions considered the resummation results obtained
in~\cite{bib-C-resummation}. For the fits to the jet broadening
measures we used the improved calculation of
Reference~\cite{bib-new-jetbroadening}.
For the sake of direct comparison with other published results we chose 
the $\ln(R)$-matching scheme\cite{bib-lnR-matching} to merge the 
\oaa\cite{bib-ERT} with the NLLA
calculations.  The renormalisation scale factor, $\xmu \equiv \mu/\sqrt{s}$
was set to $\xmu = 1$ for the main result.

The fit ranges for each observable were chosen to be the largest range for 
which the hadronisation uncertainties remained less than $15\%$, for which 
the contribution of an extreme bin did not dominate the \chisqd,
and for which fits were stable.
The remaining changes in \as\ when modifying the fit range by
one bin on either side were taken as systematic uncertainties.  Only
statistical errors were considered in the fit.

In order to investigate the importance of higher order terms in the
theory, we also changed the renormalisation scale factor in the range
of $\xmu = 0.5$ to $2.0$. The associated renormalisation scale
uncertainties are larger than those from the
detector correction and the hadronisation model dependence.

Our results for \as\ from $B_T$, $B_W$ and the $C$-parameter at
$\sqrt{s} = 35$ and $44$ GeV are given in
Table~\ref{tab-asresult-new}, where also the statistical and systematic
uncertainties of the measurements are given.
It should be pointed out that the improved perturbative calculation
for the jet broadening resulted in an increased \as\ value,
not affecting the systematic uncertainties but improving the
\chisqd

\begin{figure}[!htb]
\begin{center}
\resizebox{120mm}{!}{\includegraphics{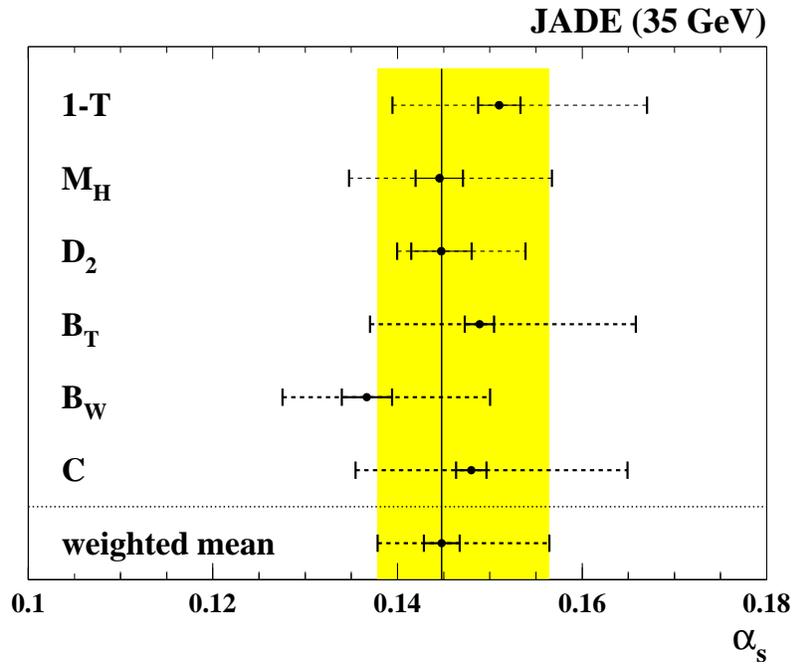}} \\
\resizebox{120mm}{!}{\includegraphics{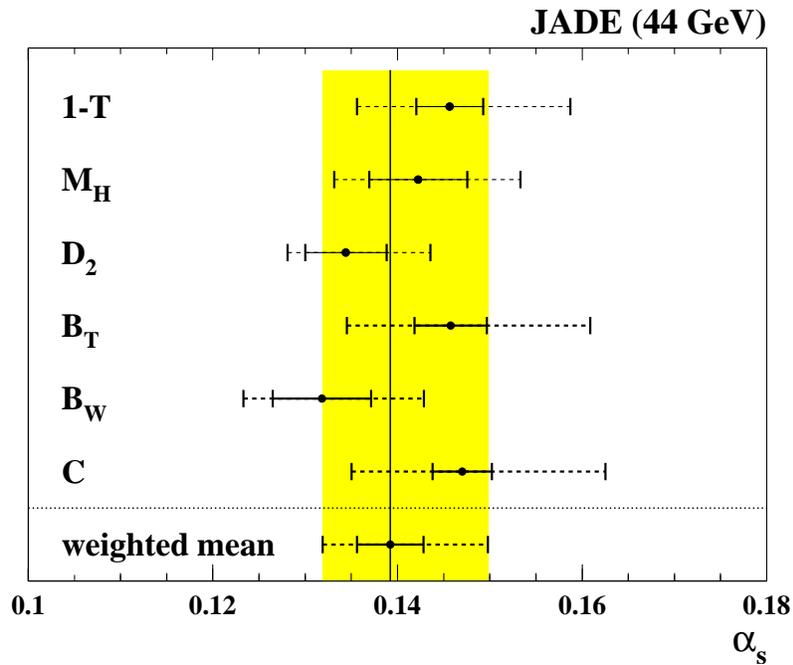}}
\end{center}
\caption{\label{fig-asresult-numbers}
Values of \as(35~GeV) and \as(44~GeV) derived from \oaa+NLLA fits to 
event shape distributions. The experimental and statistical uncertainties
are represented by the solid error bars. The dashed error bars show the
total error including hadronisation and higher order effects. The shaded
region shows the one standard deviation region around the weighted average
(see text).
}
\end{figure}

In Figure~\ref{fig-asresult-numbers} the \as\ values from this study are
shown with those obtained from the
distributions of thrust, heavy jet mass and the differential
2-jet rate using the Durham scheme as presented in our previous
publication~\cite{bib-newJADE}. 
Replacing the \as\ values obtained from the jet broadening observables,
a single \as\ value was obtained from the
individual determinations from the six event shape observables
following the procedure described in
References~\cite{bib-OPALresummed}.  
This procedure accounts for correlations of the systematic
uncertainties.  At each energy, a weighted average of the six \as\ 
values was calculated with the reciprocal of the respective squared 
total error used as a weight.  For each of the systematic
checks, the mean of the \as\ values from all considered observables
was determined.  Any deviation of this mean from the weighted average
of the main result was taken as a systematic uncertainty.

The final results for \as\ are
\begin{eqnarray*}
\as(35\ {\mathrm{GeV}}) & = & 0.1448 \pm 0.0010{\mathrm{(stat.)}}
                         \ ^{+0.0117}
                           _{-0.0069}{\mathrm{(syst.)}} \\
\as(44\ {\mathrm{GeV}}) & = & 0.1392 \pm 0.0017{\mathrm{(stat.)}}
                         \ ^{+0.0104}
                           _{-0.0072}{\mathrm{(syst.)}} \ ,
\end{eqnarray*}
where the systematic errors at $35$ and $44$~GeV are the quadratic sums of
the experimental uncertainties ($\pm 0.0017$, $\pm 0.0032$), the
effects due to the Monte Carlo modelling ($^{+0.0070}_{-0.0035}$,
$^{+0.0050}_{-0.0027}$) and the contributions due to variation of
the renormalisation scale ($^{+0.0092}_{-0.0057}$,
$^{+0.0086}_{-0.0058}$). The modelling uncertainties due to quark mass
effects contribute significantly to the total error.
  
\section{Power corrections to the mean values of event shapes }
\label{sec-meanvalues}

The strong coupling constant \as\ can also be extracted from the
energy dependence of the mean values of event shape distributions.
Non-perturbative effects are inherent in the event
shape observables due to the contributions of very low energetic
gluons and of hadronisation. For an observable ${\cal F}$ the effect
of the gluons 
can be 
represented by additive power-suppressed
corrections ($\langle {\cal F}^{\mathrm{pow.}}\rangle\propto 1/\sqrt{s}$) 
to the perturbative predictions 
of the mean values of the event shape distributions 
($\langle {\cal F}^{\mathrm{pert.}}\rangle$).  In the calculations of
Reference~\cite{bib-Milan-factor,bib-webber} which we used in this analysis 
a non-perturbative parameter
\begin{displaymath}
\bar{\alpha}_0(\mu_I) = 
   \frac{1}{\mu_I} 
   \int_0^{\mu_I} {\mathrm{d}}k\ \ \as(k)
\end{displaymath}
was introduced to replace the divergent portion of the perturbative
expression for $\as(\sqrt{s})$ below an infrared matching scale $\mu_I$. 
The power corrections to event shape observables have been
calculated in References~\cite{bib-Milan-factor} up to two-loops. 
It was found that the two-loop result is identical to the one-loop
calculation of Reference~\cite{bib-webber} up to a factor 
${\cal M} = 1.79 \pm 0.36$, which is known as the Milan factor, and
a factor $2/\pi$, which is due to defining the non-perturbative parameter 
$\bar{\alpha}_0$ as above.

The power corrections for the mean values of 
thrust ($T$), heavy jet mass ($M_H^2/s$) and $C$-parameter is 
\begin{displaymath}
\langle {\cal F}^{\mathrm{pow.}}\rangle = 
a_{\cal F} \cdot \frac{16}{3\pi} \frac{2{\cal M}}{\pi} \cdot 
\left(\frac{\mu_I}{\sqrt{s}}\right)\cdot 
   (\bar{\alpha}_o(\mu_I) - \as(\sqrt{s}) + \oaa)
\ \ \ .
\end{displaymath}
It was recently found in 
Reference~\cite{bib-new-broadening} that an additional factor has 
to be applied in the case of jet broadening observables.
In leading order the factors are given by
\begin{displaymath}
  \frac{\pi}{4\sqrt{\as(0.472\sqrt{s})/3}}  - 0.822\ \ \ ,\ \ \ %
  \frac{\pi}{4\sqrt{2\as(0.472\sqrt{s})/3}} - 0.343
\end{displaymath}
for total ($B_T$) and wide ($B_W$) jet broadening, respectively.
The $a_{\cal F}$ coefficients which depend on the observable ${\cal F}$
are calculable. They assume the values $-1$, $1/2$, $1/2$, $1/4$, $3\pi/2$ 
for thrust, heavy jet mass, total and wide jet broadening, 
and $C$-parameter, respectively~\cite{bib-Milan-factor,bib-new-broadening}.

The power correction to the mean values of an event shape ${\cal F}$
is an additive term to the perturbative prediction,
$ \langle {\cal F}\rangle = \langle {\cal F}^{\mathrm{pert.}} \rangle + 
                           \langle {\cal F}^{\mathrm{pow.}}  \rangle$. 
We determined $\as(\mz)$ and the non-perturbative parameter
$\bar{\alpha}_0(\mu_I)$ from \chisq\ fits of this 
expression to the mean values of thrust, heavy jet mass, 
$C$-parameter, total and wide jet broadening. The fits included 
the measured mean values published by various experiments at 
different centre-of-mass
energies~\cite{bib-newJADE,bib-meanvalues}
between $14$ and $183$~GeV.
For the central fit results of \as, we fixed the renormalisation scale
factor $\xmu$ to one and the infrared scale to $\mu_I=2$~GeV.
Theoretical systematic uncertainties were assessed by varying \xmu\ 
from $0.5$ to $2$, $\mu_I$ from $1$ to $3$~GeV, and ${\cal M}$ within
the quoted uncertainty.

\begin{figure}[!htb]
\vspace*{-7mm}
\begin{center}
\resizebox{78mm}{!}{\includegraphics{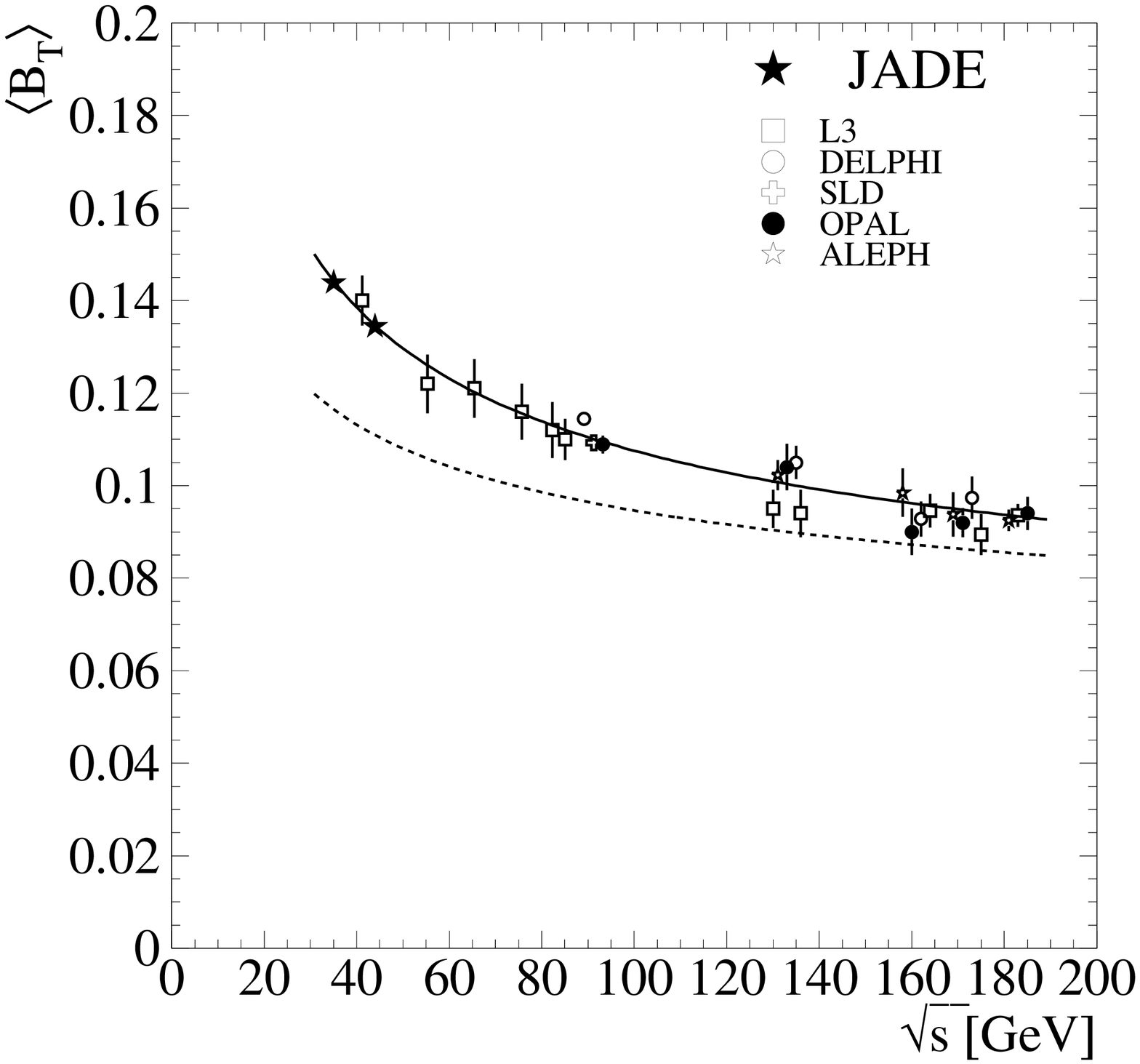}}
\resizebox{78mm}{!}{\includegraphics{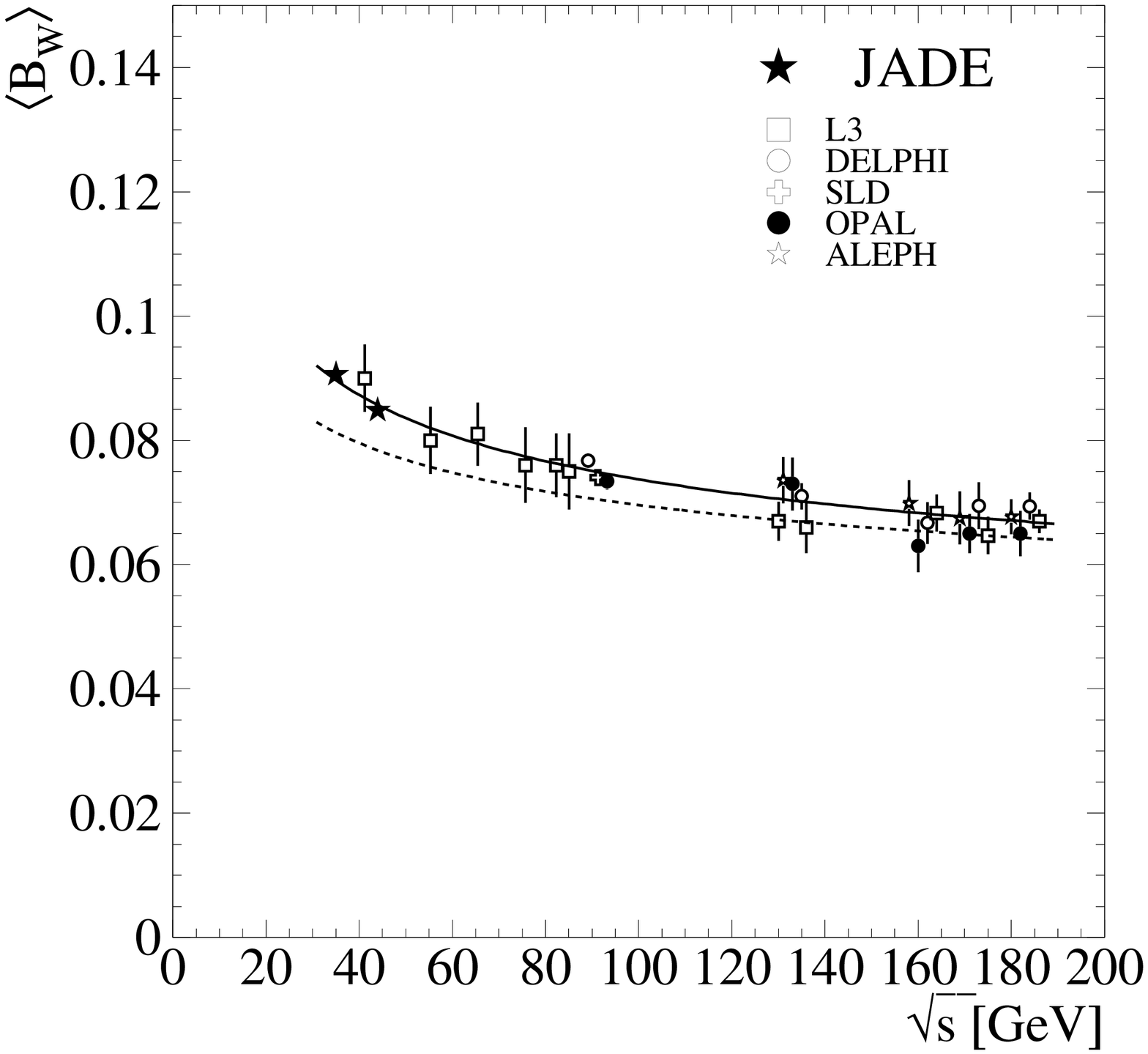}} \\[4mm]
\resizebox{78mm}{!}{\includegraphics{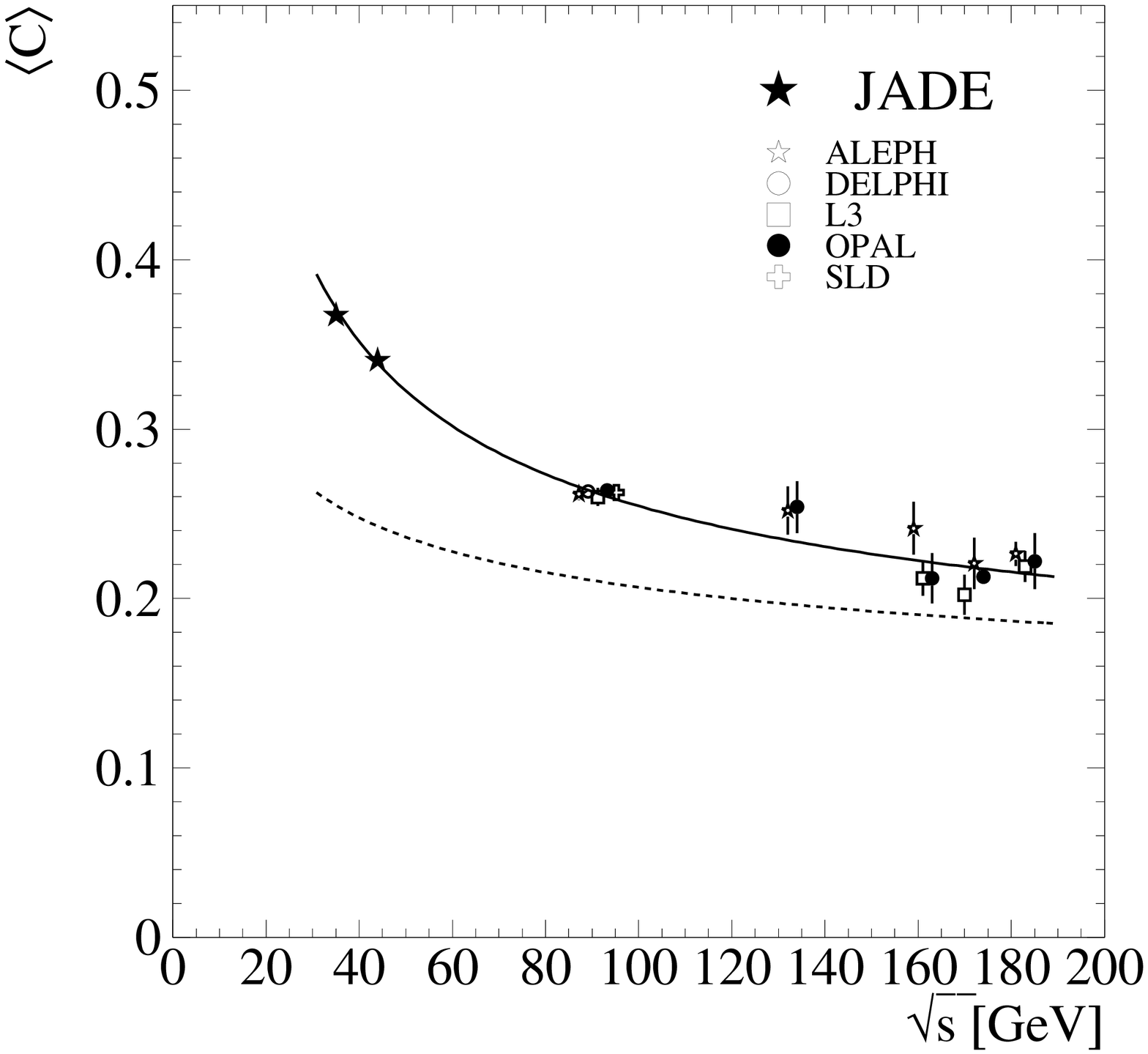}} 
\end{center}
\vspace*{-5mm}
\caption{\label{fig-as-powcor}
Energy dependence of the mean values of 
the total ($\langle B_T\rangle$) and wide jet broadening ($\langle
B_W\rangle$), and of the $C$-parameter ($\langle C\rangle$) are
shown~\protect\cite{bib-meanvalues}.
The solid curves are the result of the fit using perturbative
calculations plus two-loop power corrections which include the Milan
factor \protect\cite{bib-Milan-factor} and the revisited power
corrections to jet broadening observables\protect\cite{bib-new-broadening}.
The dashed line is the perturbative prediction using the fitted value of 
\asmz.}
\end{figure}

Figure~\ref{fig-as-powcor} shows the data values and the fit curves 
for the $C$-parameter, the total and the wide jet broadening, $B_T$ and 
$B_W$. The \chisqd\ of the fits is between
$0.7$ (for the heavy jet mass) and $1.3$ (for the thrust).
The numeric results of all fits are summarized in
Table~\ref{tab-as-powcor}, presenting the values for $\as(\mz)$ and
for $\bar{\alpha}_0(\mu_I)$, their experimental errors and systematic
uncertainties. We also quote averages of the individual \as\ and
$\bar{\alpha}_0$ results which were calculated according to the 
procedure used in Section~\ref{sec-alphas}. We consider these 
results as a test of the new theoretical prediction~\cite{bib-Milan-factor,
bib-new-broadening} for the power corrections to the mean values of 
event shapes. It should be noted that the individual results for
$\bar{\alpha}_0(\mu_I)$ scatter around the average value of 
$\bar{\alpha}_0(2 {\mathrm{~GeV}}) = 0.473 ^{+0.058}_{-0.041}$,
where the error is the total uncertainty. The r.m.s.\ of the scatter is 
$0.076$. The theoretically expected universality of $\bar{\alpha}_0$ is, 
therefore, observed at the level of better than $20\%$.

Our final value of the strong coupling from the fits to the energy
dependence of the mean values of event shapes is
\begin{displaymath}
   \as(\mz) = 0.1177\ ^{+0.0035}_{-0.0034}\ \ \ ,
\end{displaymath}
where the error is the total uncertainty.
This result lies about 2\% above the value obtained in our previous
analysis~\cite{bib-newJADE}, which did not consider the new Milan
correction nor the revisited calculation for the power correction
of the jet broadening observables. Our updated result is in good 
agreement with the world average value~\cite{bib-world-alphas-sb} 
of $\as^{\mathrm{w.a.}}(\mz)=0.119\pm 0.004$ and is of similar
precision.

\section{Summary}
\label{sec-conclusions}

Inclusion of JADE data and the availability of improved calculations 
substantially increased the significance and consistency of QCD tests
based on measurements of hadronic event shapes.

We updated our determinations~\cite{bib-newJADE} of the strong coupling 
constant \as\ at $\sqrt{s} = 35$ and $44$~GeV using $\ln(R)$-matching
for the combination of \oaa\ and resummed calculations
now including the
$C$-parameter.
Improved perturbative
calculations to the jet broadening measures $B_T$ and $B_W$ were
applied. 
We found these
calculations 
to describe the data better.
The $\as$ values were also found to be more consistent with
those from other event shape observables.
Evolving our combined \as\ measurements to $\sqrt{s} = \mz$ we
obtain $0.123\,^{+0.008}_{-0.005}$ and $0.123\,^{+0.008}_{-0.006}$
from $35$ and $44$~GeV data, respectively, resulting in a 
combined value of $\asmz = 0.123\,^{+0.008}_{-0.005}$.

We further investigated mean values of
the event shape distributions for the energy range 
of the PETRA and LEP colliders
and compared directly with analytic
QCD predictions plus power corrections for hadronisation effects, the
latter involving a universal non-perturbative parameter
$\bar{\alpha}_0$.
Our present studies, based on two-loop calculations for the event 
shapes, 
yield
$\as(\mz) = 0.1177\ ^{+0.0035}_{-0.0034}$,
which is in good agreement with our results
from the \oaa+NLLA fits and also with the world average value. The
expected universality of the non-perturbative parameter
$\bar{\alpha}_0$ is now observed at a level of better than $20$\%.

\medskip
\bigskip\bigskip\bigskip
\appendix
\section*{Acknowledgements}
\par
We are indebted to J.~Olsson for his valuable help and 
continuous effort to keep JADE data alive and 
we thank the L3 Collaboration for providing $\langle C\rangle$ data
prior to publication.

%
\clearpage

\section*{ Tables }
\begin{table}[!htb]
\begin{center}
%
%
\begin{minipage}[t]{127mm}
\begin{tabular}{|c||r@{ $\pm$ }l@{ $\pm$ }l|r@{ $\pm$ }l@{ $\pm$ }l|}   
\hline
     & \multicolumn{6}{c|}{$1/\sigma \cdot {\mathrm{d}}\sigma/{\mathrm{d}}C$} \\
\cline{2-7}
 \raisebox{1.5ex}[-1.5ex]{$C$}    
       & \multicolumn{3}{c|}{$\sqrt{s} = 35 {\mathrm{~GeV}} $} 
             &  \multicolumn{3}{c|}{$\sqrt{s} = 44 {\mathrm{~GeV}} $}  \\
\hline\hline
$ 0.00 $-$0.08  $&$  0.064  $&$   0.004  $&$   0.034 $&$   0.093 $&$     0.008 $&$     0.031   $\\
$ 0.08 $-$0.12  $&$  0.434  $&$   0.019  $&$   0.079 $&$   0.801 $&$     0.047 $&$     0.131   $\\
$ 0.12 $-$0.16  $&$  1.383  $&$   0.039  $&$   0.068 $&$   1.953 $&$     0.086 $&$     0.250   $\\
$ 0.16 $-$0.20  $&$  2.436  $&$   0.056  $&$   0.068 $&$   3.016 $&$     0.116 $&$     0.189   $\\
$ 0.20 $-$0.24  $&$  2.678  $&$   0.060  $&$   0.229 $&$   3.172 $&$     0.123 $&$     0.288   $\\
$ 0.24 $-$0.28  $&$  2.845  $&$   0.062  $&$   0.260 $&$   2.759 $&$     0.116 $&$     0.141   $\\
$ 0.28 $-$0.32  $&$  2.289  $&$   0.054  $&$   0.208 $&$   2.457 $&$     0.108 $&$     0.286   $\\
$ 0.32 $-$0.36  $&$  2.327  $&$   0.056  $&$   0.269 $&$   1.748 $&$     0.086 $&$     0.098   $\\
$ 0.36 $-$0.40  $&$  1.837  $&$   0.047  $&$   0.109 $&$   1.623 $&$     0.082 $&$     0.230   $\\
$ 0.40 $-$0.44  $&$  1.441  $&$   0.041  $&$   0.131 $&$   1.163 $&$     0.068 $&$     0.150   $\\
$ 0.44 $-$0.48  $&$  1.254  $&$   0.038  $&$   0.066 $&$   1.159 $&$     0.070 $&$     0.210   $\\
$ 0.48 $-$0.52  $&$  1.024  $&$   0.034  $&$   0.071 $&$   0.847 $&$     0.056 $&$     0.071   $\\
$ 0.52 $-$0.58  $&$  0.839  $&$   0.025  $&$   0.034 $&$   0.720 $&$     0.041 $&$     0.070   $\\
$ 0.58 $-$0.64  $&$  0.702  $&$   0.022  $&$   0.095 $&$   0.626 $&$     0.038 $&$     0.046   $\\
$ 0.64 $-$0.72  $&$  0.532  $&$   0.017  $&$   0.030 $&$   0.503 $&$     0.030 $&$     0.050   $\\
$ 0.72 $-$0.82  $&$  0.453  $&$   0.014  $&$   0.051 $&$   0.348 $&$     0.022 $&$     0.034   $\\
$ 0.82 $-$1.00  $&$  0.090  $&$   0.004  $&$   0.010 $&$   0.079 $&$     0.008 $&$     0.015   $\\
\hline\hline
mean value       &$  0.3673 $&$     0.0013 $&$    0.0040 $&$  0.3404 $&$     0.0023 $&$    0.0037 $\\
\hline
\end{tabular}
\end{minipage}
\caption[dummy]{\label{tab-eventshapes-35+44GeV}
Event shape data at $\protect\sqrt{s}=35$ (left) and $44$~GeV (right) 
for the $C$-parameter observable.
The values were corrected for detector 
and for initial state radiation effects. 
The first error denotes the statistical and the second 
the experimental systematic uncertainty.}
\end{center}
\end{table}

\begin{table}[!hb]
\vspace*{-4mm}
\begin{center}
\begin{tabular}{|c|c||c|c|c|c|c|}
\hline
 $\sqrt{s}$ & Observable  & $\as(\sqrt{s})$ & stat. &  exp. & MC &  renorm. \\
\hline\hline
         &
   $B_T$ & $0.1489$ & $\pm 0.0008$ & $\pm 0.0014$ 
         & \raisebox{-1.ex}{$\stackrel{\textstyle  +0.0099}{-0.0048}$} 
         & \raisebox{-1.ex}{$\stackrel{\textstyle  +0.0136}{-0.0107}$} \\
\cline{2-7}
 $35$ GeV&
   $B_W$ & $0.1367$ & $\pm 0.0009$ & $\pm 0.0026$
         & \raisebox{-1.ex}{$\stackrel{\textstyle  +0.0089}{-0.0045}$} 
         & \raisebox{-1.ex}{$\stackrel{\textstyle  +0.0096}{-0.0075}$} \\
\cline{2-7}
         &
    $C$  & $0.1480$ & $\pm 0.0009$ & $\pm 0.0014$ 
         & \raisebox{-1.ex}{$\stackrel{\textstyle  +0.0097}{-0.0058}$} 
         & \raisebox{-1.ex}{$\stackrel{\textstyle  +0.0138}{-0.0110}$} \\
\hline\hline
         &
   $B_T$ & $0.1458$ & $\pm 0.0014$ & $\pm 0.0037$ 
         & \raisebox{-1.ex}{$\stackrel{\textstyle  +0.0072}{-0.0032}$} 
         & \raisebox{-1.ex}{$\stackrel{\textstyle  +0.0127}{-0.0100}$} \\
\cline{2-7}
 $44$ GeV&
   $B_W$ & $0.1318$ & $\pm 0.0016$ & $\pm 0.0051$
         & \raisebox{-1.ex}{$\stackrel{\textstyle  +0.0054}{-0.0026}$} 
         & \raisebox{-1.ex}{$\stackrel{\textstyle  +0.0081}{-0.0061}$} \\
\cline{2-7}
         &
    $C$  & $0.1470$ & $\pm 0.0017$ & $\pm 0.0027$ 
         & \raisebox{-1.ex}{$\stackrel{\textstyle  +0.0073}{-0.0044}$} 
         & \raisebox{-1.ex}{$\stackrel{\textstyle  +0.0133}{-0.0107}$} \\
\hline
\end{tabular}
\end{center}
\vspace*{-4mm}
\caption{\label{tab-asresult-new}
 Values of \as\ obtained  from fits of \oaa\ +NLLA QCD calculations to the distributions
 of total and wide jet broadening, $B_T$ and $B_W$, and $C$-parameter, at 
 $\protect\sqrt{s} =~35$ and $44$~GeV. Additionally, the statistical
 (stat.) and the experimental errors (exp.) of the fit, 
 the uncertainties due to the Monte Carlo modelling (MC) of the hadronisation 
 and due to the choice of the renormalisation scale (renorm.) are given.}
\end{table}

\newpage

\begin{table}
\begin{center}
\begin{tabular}{|r||r|r|r|r|r||r|}   \hline
  \multicolumn{1}{|c||}{(a)} 
 &\multicolumn{1}{c|}{$\langle 1-T     \rangle$} 
 &\multicolumn{1}{c|}{$\langle M_H^2/s \rangle$}
 &\multicolumn{1}{c|}{$\langle B_T     \rangle$}  
 &\multicolumn{1}{c|}{$\langle B_W     \rangle$}
 &\multicolumn{1}{c||}{$\langle C       \rangle$}
 &\multicolumn{1}{|c|}{average}     \\
\hline\hline
$\alpha_S(\mz)$
    &\bf 0.1198  &\bf 0.1141  &\bf 0.1183  &\bf 0.1190  &\bf 0.1176               &\bf 0.1177 \\
\hline\hline
$Q$ range [GeV]
    & $13$-$183$ & $14$-$183$ & $35$-$183$ & $35$-$183$ & $35$-$183$              &              \\
\hline\hline
$\chi^2/{\mathrm{d.o.f.}}$
    & $52.2/39$  & $22.0/33$  & $22.1/25$  & $18.8/26$  & $18.8/16$               &              \\
\hline\hline
experimental
    &$\pm 0.0013$&$\pm 0.0010$&$\pm 0.0016$&$\pm 0.0020$&$\pm 0.0013$             &$\pm 0.0016$ \\
\hline\hline
$x_{\mu}=0.5$    
    &  $-0.0049$ &  $-0.0026$ &  $-0.0038$ &  $+0.0017$ &  $-0.0043$              &  $-0.0027$  \\
\hline
$x_{\mu}=2.0$     
    &  $+0.0061$ &  $+0.0037$ &  $+0.0048$ &  $+0.0003$ &  $+0.0053$              &  $+0.0026$  \\
\hline\hline
${\cal M}-20\%$
    &  $+0.0011$ &  $+0.0013$ &  $+0.0008$ &  $+0.0005$ &  $+0.0009$              & $+0.0008$   \\ 
\hline
${\cal M}+20\%$
    &  $-0.0011$ &  $-0.0001$ &  $-0.0007$ &  $-0.0005$ &  $-0.0009$              & $-0.0005$  \\
\hline\hline
$\mu_I=1$~GeV   
    &  $+0.0025$ &  $+0.0013$ &  $+0.0017$ &  $+0.0011$ &  $+0.0020$              & $+0.0014$   \\ 
\hline
$\mu_I=3$~GeV 
    &  $-0.0019$ &  $-0.0011$ &  $-0.0014$ &  $-0.0009$ &  $-0.0016$              & $-0.0012$  \\
\hline\hline
\raisebox{2mm}{Total error}       
    &$\stackrel{\textstyle +0.0068}{-0.0055}$ 
           &$\stackrel{\textstyle +0.0043}{-0.0030}$ 
                  &$\stackrel{\textstyle +0.0054}{-0.0044}$
                         &$\stackrel{\textstyle +0.0029}{-0.0022}$
                                &$\stackrel{\textstyle +0.0058}{-0.0049}$ 
                                              &$\stackrel{\textstyle +0.0035}{-0.0034}$ \\
\hline
\multicolumn{7}{c}{\vspace*{7mm}} \\
  \hline
  \multicolumn{1}{|c||}{(b)} 
 &\multicolumn{1}{c|}{$\langle 1-T \rangle$} 
 &\multicolumn{1}{c|}{$\langle M_H^2/s \rangle$}
 &\multicolumn{1}{c|}{$\langle B_T \rangle$}  
 &\multicolumn{1}{c|}{$\langle B_W \rangle$}
 &\multicolumn{1}{c||}{$\langle C   \rangle$} 
 &\multicolumn{1}{|c|}{average} \\
\hline\hline
$\bar{\alpha}_0(2 {\mathrm{~GeV}})$
    &\bf 0.509   &\bf 0.614   &\bf 0.442   &\bf 0.392   &\bf 0.451         & \bf 0.473  \\
   \hline\hline
experimental
    &$\pm 0.012 $&$\pm 0.018 $&$\pm 0.015 $&$\pm 0.028 $&$\pm 0.010 $      &$\pm 0.014$ \\
   \hline\hline
$x_{\mu}=0.5$    
    &  $+0.003 $ &  $+0.011 $ &  $+0.020 $ &  $+0.109 $ &  $+0.005 $       &$+ 0.018$ \\
   \hline
$x_{\mu}=2.0$     
    &  $-0.002 $ &  $-0.005 $ &  $-0.014 $ &  $-0.042 $ &  $-0.003 $       &$- 0.009$ \\
   \hline\hline
${\cal M}-20\%$    
    &  $+0.058 $ &  $+0.084 $ &  $+0.046 $ &  $+0.032 $ &  $+0.050 $       &$+ 0.053$ \\
   \hline
${\cal M}+20\%$    
    &  $-0.040 $ &  $-0.064 $ &  $-0.031 $ &  $-0.022 $ &  $-0.034 $       &$- 0.037$ \\
   \hline\hline
\raisebox{2mm}{Total error}       
    &$\stackrel{\textstyle +0.059 }{-0.042 }$
           &$\stackrel{\textstyle +0.087 }{-0.067 }$
                  &$\stackrel{\textstyle +0.052 }{-0.037 }$
                         &$\stackrel{\textstyle +0.117 }{-0.055 }$ 
                                &$\stackrel{\textstyle +0.051 }{-0.036 }$ 
                                             &$\stackrel{\textstyle +0.058 }{-0.041 }$ 
                                                        \\
   \hline
\end{tabular}
\end{center}
\caption{\label{tab-as-powcor}
Values of \asmz\ (a) and $\bar{\alpha}_0(\mu_I)$ (b) derived 
using $\mu_I=2$~GeV and $\xmu=1$ and 
the \oaa\ calculations plus two-loop power corrections, 
which include the Milan factor \protect\cite{bib-Milan-factor}, 
and the revisited power corrections for the broadening
observables~\protect\cite{bib-new-broadening}.
In addition, the
statistical and systematic uncertainties are given. Signed values 
indicate the direction in which 
\asmz\ and 
$\bar{\alpha}_0(\mu_I)$ changed 
with respect to the standard analysis. The renormalisation and infrared scale 
uncertainties are treated as an asymmetric uncertainty on 
\asmz\ and $\bar{\alpha}_0(\mu_I)$. No error contribution from the
infrared scale $\mu_I$ is assigned to $\bar{\alpha}_0(\mu_I)$.
}
\end{table}

%
\clearpage
\section*{ Figures }

\end{document}